\documentstyle{article}

\begin{document}
\large
\title{
{\LARGE\bf The non-self-adjointness of the radial momentum operator in \boldmath $n$ dimensions}}
\author{\large  Gil Paz \\ \normalsize\it Physics Department, Technion-Israel Institute of Technology, 3200 Haifa, Israel}
\date{}
\maketitle
\begin{abstract}
\large The non self-adjointness of the radial momentum operator has been noted before by several authors, but the various proofs are incorrect. We give a rigorous proof that the $n$-dimensional radial momentum operator is not self-adjoint and has no self-adjoint extensions. The main idea of the proof is to show that this operator is unitarily equivalent to the momentum operator on $L^{2}[(0,\infty),dr]$ which is not self-adjoint and has no self-adjoint extensions.     
\end{abstract}
\thispagestyle{empty}
\pagebreak
\section{INTRODUCTION}
The radial momentum operator was the subject of long discussions since the early days of Quantum Mechanics. Its exact form and  relation to the Hamiltonian were considered by many authors\cite{podolsky,dirac,dewitt,messiah}. Unlike the classical radial momentum, the connection between the radial momentum operator and the Hamiltonian of a free particle is not trivial \cite{ essen,gil}. In fact, the connection between the radial momentum and the Hamiltonian in $n$ dimensions is \cite{gil}:
\begin{displaymath}
\hat{H}=\frac{\hat{P}_{r}^{2}}{2m} + \frac{\mbox{\boldmath $ \hat{L}^2$}}{2mr^2} 
+\frac{\hbar^2}{2m}\cdot \frac{(n-1)(n-3)}{4r^{2}}.
\end{displaymath}
(at least formally, in principle one has to define the self-adjoint extension of $\hat{P}_{r}^{2}$ \cite{reed,alb}, which is not self-adjoint.) \\
Another important question that was raised is whether the radial momentum operator is an observable. Although Dirac claimed in ``The Principles of Quantum Mechanics'' that the radial momentum operator is ``real'' \cite{dirac}, many authors realized that the radial momentum operator is not self-adjoint \cite{messiah,liboff,peres,twamley}. Unfortunately none of these proofs is correct.\\
In order to be an observable the radial momentum operator should be self-adjoint. Simply checking that the eigenvalues are real (like \cite{messiah,liboff} do) is not sufficient (or necessary), one has to pay attention to the domain on which the operator is defined. Perhaps the most appealing (but incorrect) argument appears in \cite{peres,twamley} (we use units where $\hbar=1$):\\
``Since $\left[\hat{r},\hat{P}_r \right] =i$ the unitary transformation $e^{-ia\hat{P}_{r}}$ shifts the operator $\hat{r}$ by $a$ ( because $e^{ia\hat{P}_{r}}\hat{r}e^{-ia\hat{P}_{r}}=\hat{r}+a $) while leaving its spectrum invariant (being a unitary transformation). Therefore the spectrum of $\hat{r}$ must be $(-\infty,\infty)$. Since the spectrum of $\hat{r}$ is $(0,\infty)$ the operator $\hat{P}_r$ cannot be self-adjoint.''\\
This statement, had it been true, would have prevented any operator, which has canonical commutation relation with $\hat{r}$, from being self-adjoint. Unfortunately, this statement cannot be true as we can see from the following counter example.
Consider the space: $L^{2}[(0,1),dx]$, the momentum operator $\hat{P}=-i\frac{d}{dx}$ with a suitable domain is self-adjoint in that space \cite{reed}. The position operator $\hat{X}$ is a bounded self-adjoint operator. Its spectrum is of course $(0,1)$ and we have for a suitable subspace of $L^{2}[(0,1),dx]$ (say, the infinitely differentiable functions whose compact support is in (0,1)) : $\left[\hat{X},\hat{P} \right] =i$. Following the logic of the above statement we would have concluded that $\hat{P}$ is not self-adjoint. Therefore, from this simple example, we see that the above statement cannot be correct.\\
Nevertheless the operator $e^{-ia\hat{P}_{r}}$ should correspond to the translation operator on $L^2[(0,\infty),r^{n-1}dr]$. This operator is at least an isometry so it should be of the form:
\begin{equation}
e^{-ia\hat{P_{r}}}\psi(r)=\left\{ \begin{array}{ll}
                              \psi(r-a) & \mbox{if $a \leq r $}\\
  			      0  & \mbox{if $0 < r < a$}.\\    	   
     			     \end{array}                             \right.   
\end{equation}  
Such an operator would not be unitary: if $\psi(r)\neq 0$ for $r<a$ 
then the action of $e^{i\hat{P_{r}}a}$ on $\psi(r)$ is not defined. Stated differently, ``we can move everything to the right but not to the left''. Since the translation operator is not unitary we have every reason to suspect that $\hat{P}_{r}$ is not self-adjoint. Therefore, it seems that a correct proof to the non self-adjointness of the radial momentum operator is highly in order.\\
There's seem to be almost a consensus in the literature that the operator which correspond to the radial momentum in $n$ dimensions is the operator $-i \left( \frac{\partial}{\partial r}+\frac{n-1}{2r} \right)$ \cite{dewitt,essen}.
We shall now show that this operator is  not self-adjoint and has no self-adjoint extensions. 

\pagebreak
\section{THE PROOF}
It is well known that the momentum operator in $L^{2}[(0,\infty),dr]$ is not self-adjoint and has no self-adjoint extensions \cite{cek}. When we say the ``momentum operator'' we mean the operator $\hat{P}=-i\frac{d}{dr}$ with the domain: \[\{\psi | \psi\in L^{2},\psi'\in L^{2}, \psi \mbox{ is absolutely continuous on }(0,\infty),\psi(0)=0\}. \] These conditions are needed to assure that $\hat{P}$
would be symmetric (notice that since $\psi$ is absolutely continuous $\lim_{r \rightarrow\infty}\psi=0$ \cite{cek}). We are going to use this fact to prove our assertion.\\
First, define a transformation:
\begin{eqnarray}
U &:& L^{2}[(0,\infty),dr] \rightarrow   L^{2}[(0,\infty),r^{n-1}dr ] \nonumber\\
(U\psi)(r)&:=&\frac{\psi}{r^{\frac{n-1}{2}}}.
\end{eqnarray}
We have
\begin{equation}
\|\psi\|^{2}=\int_{0}^{\infty}\left|\psi\right|^{2}dr=
\int_{0}^{\infty}\left|\frac{\psi}{r^{\frac{n-1}{2}}}\right|^{2}r^{n-1}dr=
\|U\psi\|^{2},
\end{equation}
so $U$ is an isometry. $U$ is a unitary operator since every $\varphi\in L^{2}[(0,\infty),r^{n-1}dr ]$ has an inverse image $r^{\frac{n-1}{2}}\varphi\in L^{2}[(0,\infty),dr]$. This transformation (with $n=3$) is well known from elementary textbooks on quantum mechanics, where it is used to solve the Schr\"{o}dinger equation for the hydrogen atom \cite{cohen}. \\
$U^{-1}$ is defined by:
\begin{eqnarray}
U^{-1} &:& L^{2}[(0,\infty),r^{n-1}dr] \rightarrow L^{2}[(0,\infty),dr ] \nonumber\\
(U^{-1}\varphi)(r)&:=&r^{\frac{n-1}{2}}\varphi.
\end{eqnarray}
The operator $\hat{P}$ on $L^{2}[(0,\infty),dr]$ is unitarily equivalent to:
\begin{equation}
\hat{P_{r}} \stackrel{def}{=} U\hat{P}U^{-1}=r^{\frac{1-n}{2}}\left(-i\frac{d}{dr}\right)r^{\frac{n-1}{2}}=
-i\left(\frac{d}{dr}+\frac{n-1}{2r}\right),
\end{equation}
which is, formally, the n-dimensional radial momentum operator.\\
If $\varphi\in D(\hat{P}_{r})$ then $U^{-1}\varphi\in D(\hat{P})$. Therefore if  $\varphi\in D(\hat{P}_{r})$ then:
\begin{enumerate}
\item[a.] $\left(r^{\frac{n-1}{2}}\varphi\right)'\in L^{2}[(0,\infty),dr]$
\item[b.] $r^{\frac{n-1}{2}}\varphi$ is absolutely continuous in $(0,\infty)$
\item[c.] $(r^{\frac{n-1}{2}}\varphi)(0)=0$.
\end{enumerate}
The obvious question arises: is this the ``natural'' domain for $\hat{P}_{r}$ ?\\\\
First of all, $\hat{P}_{r}\varphi\in L^{2}[(0,\infty),r^{n-1}dr]$, that is:
\begin{eqnarray}
\int_{0}^{\infty}\left|\hat{P}_{r}\varphi\right|^{2}r^{n-1}dr&=&
\int_{0}^{\infty}\left|\frac{d}{dr}r^{\frac{n-1}{2}}\varphi\right|^{2}
\left(r^{\frac{1-n}{2}}\right)^{2}r^{n-1}dr \nonumber \\
&\Rightarrow&\left(r^{\frac{n-1}{2}}\varphi\right)'\in L^{2}[(0,\infty),dr],
\end{eqnarray}
and we have (a).\\
$\hat{P}_{r}$ is (at least) symmetric, that is if $\varphi,\chi\in D(\hat{P}_{r})$ then the following equality should hold:
\begin{equation}
\label{int}
\int_{0}^{\infty}\overline{\chi}\left(\hat{P}_{r}\varphi\right)r^{n-1}dr
 =
\int_{0}^{\infty}\overline{\left(\hat{P}_{r}\chi\right)}\varphi r^{n-1}dr.
\end{equation}
However,
\begin{eqnarray}
\label{intres}
\int_{0}^{\infty}\overline{\chi}\left(\hat{P}_{r}\varphi\right)r^{n-1}dr&=&
(-i)\int_{0}^{\infty}\overline{\chi}r^{\frac{1-n}{2}}\frac{d}{dr}\left(r^{\frac{n-1}{2}}\varphi\right)r^{n-1}dr\nonumber \\
&=&(-i)\int_{0}^{\infty}\overline{\left( r^{\frac{n-1}{2}}\chi\right)} \frac{d}{dr}\left( r^{\frac{n-1}{2}}\varphi\right)dr= \nonumber \\
(-i)\overline{\left(r^{\frac{n-1}{2}}\chi\right)}r^{\frac{n-1}{2}}\varphi\mid_{0}^{\infty}&+&i
\int_{0}^{\infty}\overline{\frac{d}{dr}\left( r^{\frac{n-1}{2}}\chi\right)}
r^{\frac{n-1}{2}}\varphi dr= \nonumber \\
(-i)\overline{\left(r^{\frac{n-1}{2}}\chi\right)}r^{\frac{n-1}{2}}\varphi\mid_{0}^{\infty}&+&\int_{0}^{\infty}\overline{\left(\hat{P}_{r}\chi\right)}\varphi r^{n-1}dr,
\end{eqnarray}
where we were forced to assume (b) in order to use integration by parts. (b) also ensures that  the boundary term in (\ref{intres}) is zero at infinity. In order that (\ref{int}) will hold, we have to assume that the boundary term also vanishes at the origin, i.e. assume (c).

Thus we have defined $\hat{P}_{r}$ with its proper domain and we see that it is symmetric. Furthermore it is also closed since it is unitarily equivalent to $\hat{P}$, which is closed \cite{cek}. As we have said earlier this operator is the radial momentum operator.

In order to find out whether this operator is self-adjoint or at least have self-adjoint extensions we have to check the dimensionality of the two subspaces:
${\cal K}_{-}=\ker(i+\hat{P}^{*}_{r})$ and ${\cal K}_{+}=\ker(i-\hat{P}^{*}_{r})$. If they do not have the same dimensionality the operator is not self-adjoint and has no self-adjoint extensions \cite{reed}.

$\hat{P}^{*}_{r}$ is easy to find since \cite{cek} $\hat{P}^{*}_{r}=U\hat{P}^{*}U^{-1}$  and we have:
\begin{eqnarray}
\varphi\in \ker \left(i\pm\hat{P}^{*}_{r}\right)&\Rightarrow&
\left(i\pm\hat{P}^{*}_{r}\right)\varphi=0 \Rightarrow\nonumber \\
\left(i\pm U\hat{P}^{*}U^{-1}\right)\varphi=0&\Rightarrow&
U\left(i\pm\hat{P}^{*}\right)U^{-1}\varphi=0 \Rightarrow\nonumber \\ 
\left(i\pm\hat{P}^{*}\right)U^{-1}\varphi=0&\Rightarrow&
U^{-1}\varphi\in \ker\left(i\pm\hat{P}^{*}\right).
\end{eqnarray}
In a similar way we can show that:
\begin{equation}
\psi\in \ker\left(i\pm\hat{P}^{*}\right)\Rightarrow U\psi\in \ker \left(i\pm\hat{P}^{*}_{r}\right)
\end{equation}
and we conclude that:
\begin{equation}
\dim\ker \left(i\pm\hat{P}^{*}_{r}\right)=
 \dim\ker\left(i\pm\hat{P}^{*}\right).
\end{equation}
$\hat{P}^{*}$ has the following property \cite{cek}  :
\begin{eqnarray}
\dim\ker\left(i+\hat{P}^{*}\right)&=&0 \nonumber\\
\dim\ker\left(i-\hat{P}^{*}\right)&=&1,
\end{eqnarray}
because 
\begin{eqnarray}
\ker\left(i+\hat{P}^{*}\right)&=&\{ce^{r}|c\in {\bf C} \}
\not\subseteq L^{2}[(0,\infty),dr]\nonumber\\
\ker\left(i-\hat{P}^{*}\right)&=&\{ce^{-r}|c\in {\bf C}\} \subseteq L^{2}[(0,\infty),dr].
\end{eqnarray}
Therefore
\begin{equation}
\dim\ker \left(i+\hat{P}^{*}_{r}\right)\neq \dim\ker \left(i-\hat{P}^{*}_{r}\right)
\end{equation}
 and $\hat{P}_{r}$ is not self-adjoint and does not have self-adjoint extensions.\\\\
Using this equivalence we can understand the non self-adjoint nature of the radial momentum operator on a more intuitive level, by  transferring the problem into a one dimensional problem .\\
In one dimension we can, in general, consider three types of interval: infinite interval, finite interval (``particle in a box'') and semi-infinite interval. In the first case the momentum operator is self-adjoint because we can translate a wave packet to both sides. In the case of a ``particle in a box'' we can translate a wave packet and whatever ``comes out'' at one end we can enter at the other end (possibly with a different phase which corresponds to a specific self-adjoint extension \cite{reed}). Therefore the momentum operator, which is not self-adjoint, has self-adjoint extensions. In the case of a semi-infinite interval, we can move a wave packet to the right but if we try to move it to the left what ``comes out'' at the origin cannot be entered at the other end since the ``other end'' is infinity. Therefore the translation operator is not unitary and the momentum operator is not self-adjoint and has no self-adjoint extensions.\\
To summarize, we have seen that the radial momentum operator is unitarily equivalent to the momentum operator on the half line $(0,\infty)$. Since this operator is not self-adjoint (and has no self-adjoint extensions), the radial momentum operator is not self-adjoint (and has no self-adjoint extensions).\\ \\
{\Large \bf ACKNOWLEDGMENTS}\\\\
This work was supported by the Technion Graduate School.\\\\
I thank Aviv Censor and Amnon Harel for useful discussions.\\
\pagebreak

\end{document}